# Review of Muon-Proton Collider Proposals: Main Parameters

Burak Dagli[1], Bora Ketenoglu[2,*] and Saleh Sultansoy[1,3]

[1] Department of Materials Science and Nanotechnology Engineering, TOBB University of Economics and Technology, Ankara, Turkey
[2] Department of Engineering Physics, Ankara University, Ankara, Turkey
[3] ANAS Institute of Physics, Baku, Azerbaijan

*Corresponding Author: bketen@eng.ankara.edu.tr

**Abstract**

Construction of future Muon Collider (or dedicated µ-ring) tangential to the energy frontier pp colliders will give opportunity to realize µp collisions at multi-TeV center of mass energies at a luminosity of order of $10^{33}$ cm$^{-2}$s$^{-1}$ ($10^{34}$ cm$^{-2}$s$^{-1}$). Obviously, such colliders will essentially enlarge the physics search potential of corresponding muon and hadron colliders for both the SM (especially for clarifying QCD basics) and BSM phenomena. This paper is devoted to review of main parameters of µp colliders proposed until now.

## 1. Introduction

Lepton-hadron collisions play a crucial role in our understanding of matter's structure:
- proton form-factors were first observed in electron scattering experiments [1,2]
- quarks were first observed at SLAC deep inelastic electron scattering experiments [3,4]
- EMC effect was observed at CERN in deep inelastic muon scattering experiments [5] and so on.

HERA, the first electron-proton collider, further explored structure of protons and provided parton distribution functions (PDFs) for the LHC and Tevatron. TeV (or multi-TeV) scale lepton-hadron colliders are crucial for clarifying basics of QCD, which is responsible for 98% of mass of the visible part of our Universe. It is worth to note that the electro-weak part of the Standard Model (SM) has been completed with discovery of Higgs boson at the LHC, but this is not the case for QCD part: the confinement hypothesis should be clarified. Let us mention that the confinement may be partial and free (heavy) quarks may become observable at sufficiently high energies [6,7]. Since observation of this phenomenon at pp colliders is very difficult, $l^+l^-$ and lp colliders are advantageous in this aspect. It should be emphasized that future lepton-hadron colliders will give opportunity to shed light on quark → hadron → nucleus transitions as well.

Besides, contruction of TeV energy lepton-hadron colliders is mandatory to provide PDFs for adequate interpretation of forthcoming data from HL/HE-LHC [8,9] and FCC/SppC [10,11]. Finally, energy-frontier lepton-hadron colliders are advantageous for investigation of a number of BSM phenomena (see Figure 1 for LHC, ILC and ILC*LHC comparison [12]).

Today, linac-ring type *ep* colliders are considered as sole realistic way to (multi-) TeV scale in lepton-hadron collisions (see review [13] and references therein) and LHeC [14] is the most promising candidate. However, situation may change in the coming years: µp colliders can come forward depending on progress in muon beam production and cooling issues. Let us note that muon colliders (MC) were proposed more than 50 years ago (see [15] and references therein). For recent status see Ref. [16,17].

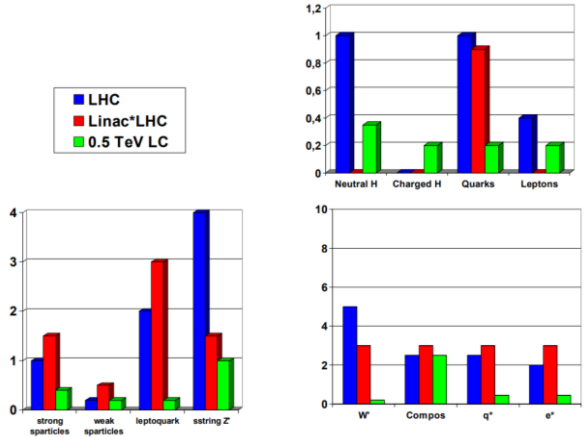

**Figure 1.** "Fingertip" estimations of discovery limits at the LHC (blue), ILC*LHC (red) and ILC (green). Upper-right picture contains: the neutral Higgs, a charged Higgs, the fourth SM family quarks and leptons. Down-left picture contains: strong sparticles (gluino and squarks), weak sparticles (neutralino, chargino and sleptons), leptoquark and Z' predicted by E6 GUT. Down-right picture contains: W', compositeness scale, excited quarks and leptons.

In Table 1, we present correlation between colliding beams and colliding schemes in energy-frontier aspect.

**Table 1.** Energy frontier colliders: colliding beams vs. collider types.

| Colliders | Ring-Ring | Linac-Linac | Linac-Ring |
|---|---|---|---|
| Hadron | + | | |
| Lepton ($e^-e^+$) | | + | |
| Lepton ($\mu^-\mu^+$) | + | | |
| Lepton-hadron (eh) | | | + |
| Lepton-hadron ($\mu$h) | + | | |
| Photon-hadron | | | + |

Muon-proton colliders [18–22] were proposed two decades ago as alternatives to linac-HERA and linac-LHC based $ep/\gamma p$ colliders (see review [23] and references therein). Two years later an ultimate $\sqrt{s}$=100 TeV µp collider (with additional 50 TeV proton ring in $\sqrt{s}$=100 TeV muon collider tunnel) was suggested in [24]. It should be noted that luminosity of µp collisions in [20] were over-estimated by 3 orders of magnitudes, respectively (see subsection 3.2 in [24]). Several years ago, FCC and SppC based energy frontier muon-hadron colliders have been proposed in [25,26] and [27], respectively. Recently, several options for LHC/FCC-based µp colliders have been considered in [28], µp colliders at the BNL have been proposed in [29], HL-LHC and HE-LHC based µp colliders have been considered in [30,31].

Chronological evolution of the smallest observable distance are given in Figure 2, where conducted experiments are marked with red dots and green dots correspond to future lepton-hadron collider proposals. The abbreviations PWFA-LC stands for $\sqrt{s}$ = 10 TeV Plasma-Wakefield Accelerator Linear Collider [32] and MC stands for $\sqrt{s}$ = 6 TeV Muon Collider [16].

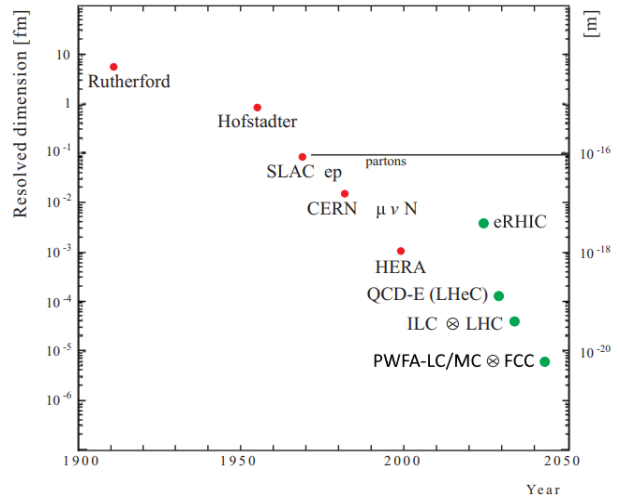

**Figure 2.** The development of the resolution power of the experiments exploring the inner structure of matter over time from Rutherford experiment to PWFA-LC/MC ⊗ FCC.

This paper is devoted to review of µp collider proposals. Here, we discuss main parameters of µp colliders. Their physics search potential as well as µA colliders will be reviewed in forthcoming papers. In section 2, we present different muon-proton collider proposals in chronological order. Parameters of these colliders are checked by the software AloHEP [33–35] that is developed for quick estimation of luminosities, tune shifts and disruptions for different types of colliders. Finally, in Section 3 we present our conclusions and recommendations.

## 2. Muon-proton collider proposals

In this section, we present different muon-proton collider proposals in chronological order. Parameters of these colliders are checked by the software AloHEP (A Luminosity Optimizer for High Energy Physics [33–35]), a luminosity estimator for ring-ring, linac-ring and linear colliders, which also computes IP parameters such as beam-beam tune shift, disruption etc.

**2.1. V. Shiltsev**, "An asymmetric µp collider as a quark structure microscope: Luminosity consideration" and "An asymmetric muon proton collider: luminosity consideration"

This is the first µp collider proposal [18,19] that considers collision of 200 GeV muons with 1 TeV protons of Tevatron, which correspond to $\sqrt{s}_{\mu p}$ = 0.89 TeV. Regarding luminosity, two options were considered: L=1.3×10$^{33}$ cm$^{-2}$s$^{-1}$ for high µ-production rate and L=1.1×10$^{32}$ cm$^{-2}$s$^{-1}$ for low µ-production rate. Main parameters of this proposal are given in Table 2.
2

**Table 2.** Two Options of *μp* Collider

| Parameter [Unit] | High μ-production | Low μ-production |
|---|---|---|
| Muon Energy [GeV] | 200 | 200 |
| Proton Energy [GeV] | 1000 | 1000 |
| # of muons per bunch [$10^{10}$] | 200 | 50 |
| Muon norm. emitt. [μm] | 50 | 200 |
| Storage turns | ~2000 | ~2000 |
| Muon pulse rate | 30 | 10 |
| Muons $\beta$ @ IP [cm] | 7.5 | 7.5 |
| Protons $\beta$ @ IP [cm] | 15 | 15 |
| Proton norm. emitt. [μm] | 12.5 | 50 |
| # of protons per bunch [$10^{10}$] | 125 | 500 |
| Proton tuneshift | 0.02 | 0.0005 |
| Muon tuneshift | 0.026 | 0.026 |
| Luminosity [cm$^{-2}$s$^{-1}$] | 1.3×10$^{33}$ | 1.1×10$^{32}$ |

Implementation of these parameters into the software AloHEP results in following luminosity values: L=1.2×10$^{33}$ cm$^{-2}$s$^{-1}$ for high μ-rate and L=1.0×10$^{32}$ cm$^{-2}$s$^{-1}$ for low μ-rate without considering losses resulting from muon decay. However, taking muon decays into account, AloHEP leads to L=0.73×10$^{33}$ cm$^{-2}$s$^{-1}$ for high μ-rate and L=0.6×10$^{32}$ cm$^{-2}$s$^{-1}$ for low μ-rate. Concerning beam-beam tune shift, AloHEP gives $\xi_\mu$=0.027 for both options, $\xi_p$=0.02 for high μ-rate and $\xi_p$=0.0012 for low μ-rate. The last value is 2.4 times larger than the value given in Table 2, which is not so crucial.

**2.2. I. F. Ginzburg**, "Physics at future *ep*, *γp* (linac-ring) and *μp* colliders"

Second time *μp* colliders have been considered in [20]. Herein, only center-of-mass energy and luminosity values have been provided referring to private communication with A. N. Skrinsky: √s$_{μp}$ = 4 TeV and L = 3×10$^{35}$ cm$^{-2}$s$^{-1}$. As mentioned in [24], luminosity value is overestimated by 2 orders of magnitude. Therefore, real luminosity of proposed *μp* collider is comparable with that of linac-ring type *ep* colliders with the same energy range, and hence, their physics search potentials are similar in contrast with the statement given in [20].

**2.3. S. Sultansoy**, "The post-HERA era: brief review of future lepton-hadron and photon-hadron colliders"

This is a most speculative (however, very attractive) one among the lepton-hadron collider options considered in review [24]. It assumes construction of additional 50 TeV proton ring in a tunnel of √s = 100 TeV $\mu^+\mu^-$ collider with L$_{μμ}$ = 10$^{36}$ cm$^{-2}$s$^{-1}$. Parameters of muon and proton beams are given in Table 3 (see Table VI and subsection 4.3 in [24]).

The luminosity of μp collisions was estimated by using equation below [24]:

$$L_{\mu p} = \frac{n_p}{n_\mu} \cdot \frac{\beta_\mu^*}{\beta_p^*} \cdot \frac{m_\mu}{m_p} \cdot \frac{\varepsilon_\mu^N}{\varepsilon_p^N} \cdot L_{\mu\mu} \quad (1)$$

resulting in L$_{μp}$ = 10$^{33}$ cm$^{-2}$s$^{-1}$.

**Table 3.** Parameters of √s = 100 TeV *μp* Collider

| Parameter [Unit] | Muon | Proton |
|---|---|---|
| Energy [TeV] | 50 | 50 |
| # of particles per bunch [$10^{10}$] | 80 | 80 |
| Number of bunches per ring | 1 | 1 |
| Circumference [km] | 100 | 100 |
| Repetition rate [Hz] | 7.9 | - |
| Number of turns | 1350 | - |
| Norm. emitt. [μm] | 8.7 | 30 |
| $\beta$ @ IP [cm] | 0.25 | 10 |

Implementation of parameters from Table 3 into the software AloHEP results in L=0.96×10$^{33}$ cm$^{-2}$s$^{-1}$ without considering losses resulting from muon decay. However, taking muon decays into account, AloHEP leads to L=0.78×10$^{33}$ cm$^{-2}$s$^{-1}$. Concerning beam-beam tune shift, AloHEP gives $\xi_\mu$=0.1 and $\xi_p$=0.003.

While there is no problem with $\xi_p$, $\xi_\mu$=0.1 is unacceptably high. As an alternative option let us use parameters of proton beam of the ERL60 upgraded FCC [26]. In this case AloHEP predicts L=1.1×10$^{33}$ cm$^{-2}$s$^{-1}$ (0.89×10$^{33}$ cm$^{-2}$s$^{-1}$) without (with) muon decay losses, $\xi_\mu$=0.012 and $\xi_p$=0.044.

**2.4. K. Cheung**, "Muon-proton Colliders: Leptoquarks, Contact Interactions and Extra Dimensions"

The center-of-mass energies and luminosities for various *μp* colliders considered in [22] are presented in Table 4.

**Table 4.** The center-of-mass energies √s and luminosities *L* for various designs of muon-proton colliders.

| Energies of muon and proton | $\sqrt{s}$ [GeV] | $L$ [$fb^{-1}$] |
|---|---|---|
| 30 GeV ⊗ 820 GeV | 314 | 0.1 |
| 50 GeV ⊗ 1 TeV | 447 | 2 |
| 200 GeV ⊗ 1 TeV | 894 | 13 |
| 1 TeV ⊗ 1 TeV | 2000 | 110 |
| 2 TeV ⊗ 3 TeV | 4899 | 280 |

The 200 GeV ⊗ 1 TeV option was proposed by Shiltsev four years before [22] and has been discussed in Section 2.1. Concerning other options, luminosity values can not be checked by AloHEP since author did not provide muon and proton beams parameter sets.

**2.5. Y. C. Acar et al.**, "FCC based *ep* and *μp* colliders"

After 15 years, muon-proton colliders have been reintroduced in [25] within the context of the FCC. Possible layout is shown in Figure 2.

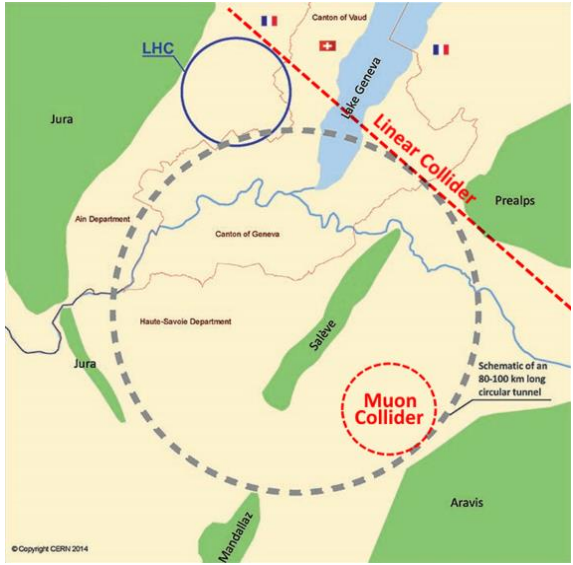

**Figure 3.** Possible configuration for FCC, linear collider (LC) and muon collider (MC).

Parameters of proton and muon beams used for estimations of μp luminosities are presented in Tables 5 and 6, respectively.

**Table 5.** Main parameters of the FCC pp option.

| Beam Energy [TeV] | 50 |
|---|---|
| Peak Luminosity [$10^{34}$ cm$^{-2}$s$^{-1}$] | 5 |
| Particle per Bunch [$10^{10}$] | 10 |
| Normalized Emittance [μm] | 2.2 |
| $\beta^*$ at IP [cm] | 110-30 |
| IP beam size [μm] | 6.8 |
| Bunches per beam | 10600 |
| Bunch spacing [ns] | 25 |
| Bunch length [mm] | 80 |

**Table 6.** Muon collider parameters

| $\sqrt{s}$ [TeV] | 0.126 | 0.35 | 1.5 | 3.0 | 6.0 |
|---|---|---|---|---|---|
| L [$10^{34}$ cm$^{-2}$s$^{-1}$] | 0.008 | 0.6 | 1.25 | 4.4 | 12 |
| Circumference [km] | 0.3 | 0.7 | 2.5 | 4.5 | 6 |
| Repetition rate [Hz] | 15 | 15 | 15 | 12 | 6 |
| $\beta^*$ at IP [cm] | 1.7 | 0.5 | 1 | 0.5 | 0.25 |
| Muon/Bunch [$10^{12}$] | 4 | 3 | 2 | 2 | 2 |
| Bunches per beam | 1 | 1 | 1 | 1 | 1 |
| $\varepsilon_N$ [π mm·rad] | 0.2 | 0.05 | 0.025 | 0.025 | 0.025 |

Luminosity values given in Table 7 are derived from the following equation [25]:

$$L_{\mu p} = \left(\frac{n_p}{n_\mu}\right)\left(\frac{\sigma_\mu}{\max[\sigma_p, \sigma_\mu]}\right)^2 L_{\mu\mu} \qquad (2)$$

**Table 7.** Main Parameters of the FCC based *μp* colliders

| Collider name | $E_\mu$ [TeV] | $\sqrt{s}$ [Tev] | $L_{\mu p}$ [$10^{31}$ cm$^{-2}$s$^{-1}$] |
|---|---|---|---|
| μ63-FCC | 0.063 | 3.50 | 0.2 |
| μ175-FCC | 0.175 | 5.92 | 20 |
| μ750-FCC | 0.75 | 12.2 | 50 |
| μ1500-FCC | 1.5 | 17.3 | 50 |
| μ3000-FCC | 3 | 24.5 | 300 |

Implementing parameters from Tables 5 and 6 into AloHEP, we obtain values of luminosity and beam-beam tune shifts given in Table 8, where muon decays have been involved. We did not include μ63-FCC and μ175-FCC since unacceptable asymmetry of muon and proton beam energies.

**Table 8.** AloHEP results for the FCC based *μp* colliders

| Collider name | $L_{\mu p}$ [$10^{31}$cm$^{-2}$s$^{-1}$] | $\xi_p$ | $\xi_\mu$ |
|---|---|---|---|
| μ750-FCC | 62 | 0.11 | 0.0043 |
| μ1500-FCC | 55 | 0.11 | 0.0043 |
| μ3000-FCC | 41 | 0.11 | 0.0043 |

It is seen from Table 8 that proton beam-beam tune shift is unacceptably high. A way to reduce $\xi_p$ down to 0.01 is reducing $N_\mu$ by an order of magnitude which results in corresponding decrement of luminosity value. In principle, this can be compensated by increment of $N_p$. It seems quite possible since low number of proton bunches are needed for *μp* colliders. For instance, 20 proton bunches are sufficient for μ750-FCC.

**2.6. Y. C. Acar et al.**, "Future circular collider based lepton–hadron and photon–hadron colliders: Luminosity and physics"

Reference [26] is an advanced version of reference [25]. Parameters of muon beams used for estimations of *μp* luminosities are presented in Table 9, parameters of FCC proton beam are given in Table 5 of previous section.

**Table 9.** Muon collider parameters

| $\sqrt{s}$ [TeV] | 0.126 | 1.5 | 3.0 |
|---|---|---|---|
| L [$10^{34}$ cm$^{-2}$s$^{-1}$] | 0.008 | 1.25 | 4.4 |
| Circumference [km] | 0.3 | 2.5 | 4.5 |
| Repetition rate [Hz] | 15 | 15 | 12 |
| $\beta^*$ at IP [cm] | 1.7 | 1 | 0.5 |
| Muon/Bunch [$10^{12}$] | 4 | 2 | 2 |
| Bunches per beam | 1 | 1 | 1 |
| $\varepsilon_N$ [π mm·rad] | 0.2 | 0.025 | 0.025 |
| Bunch length [cm] | 6.3 | 1 | 0.5 |
| Beam size at IP [μm] | 75 | 6 | 3 |
| $\xi_\mu$ at IP | 0.02 | 0.09 | 0.09 |

Using Equation 2, one obtains μp collider parameters given in Table 10 (see [26] for details).

**Table 10.** Main parameters of the FCC based *μp* colliders

| Collider name | $\sqrt{s}$ [TeV] | $L_{\mu p}$ [$10^{31}$ cm$^{-2}$s$^{-1}$] | $\xi_p$ [$10^{-2}$] | $\xi_\mu$ [$10^{-4}$] |
|---|---|---|---|---|
| μ63-FCC | 3.50 | 0.20 | 0.18 | 5.4 |
| μ750-FCC | 12.2 | 49 | 11 | 33 |
| μ1500-FCC | 17.3 | 43 | 11 | 8.3 |

Implementing parameters from Tables 5 and 9 into AloHEP, we obtain values of luminosity and beam-beam tune shifts given in Table 11. It is seen that luminosity values by AloHEP are slightly higher than the luminosities calculated using Equation 2. The reason for this is that $L_{\mu\mu}$ in Equation 2 include decays for both muon beams, whereas there is only one muon beam in *μp* colliders.

**Table 11.** AloHEP results for the FCC based *μp* colliders

| Collider name | $L_{\mu p}$ [$10^{31}$cm$^{-2}$s$^{-1}$] | $\xi_p$ | $\xi_\mu$ |
|---|---|---|---|
| μ750-FCC | 62 | 0.11 | 0.0043 |
| μ1500-FCC | 55 | 0.11 | 0.0043 |

As mentioned in previous section, proton beam-beam tune shift value is so high, and this problem can be solved by increasing of muon bunch population. Another solution was considered in [26]. Proton and muon beam tune shifts are given by:

$$\xi_p = \frac{N_\mu r_p \beta_p^*}{4\pi \gamma_p \sigma_\mu^2} \quad (3)$$

$$\xi_\mu = \frac{N_p r_\mu \beta_\mu^*}{4\pi \gamma_\mu \sigma_p^2} \quad (4)$$

According to Equation (3), reducing of $\xi_p$ can be succeeded by decreasing of $\beta_p^*$ and/or increasing of $\sigma_\mu$. For example, decreasing $\beta_p^*$ from 1.1 m to 0.1 m (as in the upgraded option of proton beams considered in Section 2 of reference [26]) seems to solve this problem. Luminosity values presented in Table 11 assume simultaneous operation with *pp* collider. These values can be increased by an order using dedicated proton beam with larger bunch population [24]. Finally, luminosity values exceeding $10^{33}$cm$^{-2}$s$^{-1}$ can be achieved at the FCC based *μp* colliders.

**2.7. A. C. Canbay et al.**, "SppC based energy frontier lepton-proton colliders: luminosity and physics"

SppC based *μp* colliders were proposed in Reference [27]. Possible layout is presented in Figure 4.

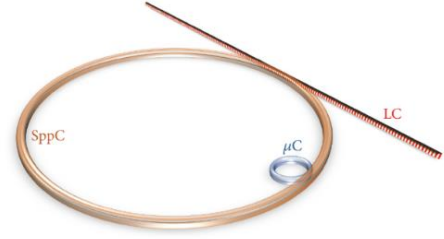

**Figure 4.** Possible configuration for SppC, linear collider (LC), and muon collider (μC).

Main parameters of SppC proton collider are given in Table 12. Muon beam parameters were given in Table 9.

**Table 12.** Main parameters of proton beams in SppC.

| Beam Energy [TeV] | 35.6 | 68.0 |
|---|---|---|
| Circumference [km] | 54.7 | 100.0 |
| Peak Luminosity [$10^{34}$ cm$^{-2}$s$^{-1}$] | 11 | 102 |
| Particle per Bunch [$10^{10}$] | 20 | 20 |
| Normalized Emittance [μm] | 4.10 | 3.05 |
| $\beta^*$ function at IP [cm] | 75 | 24 |
| IP beam size [μm] | 9.0 | 3.04 |
| Bunches per beam | 5835 | 10667 |
| Bunch spacing [ns] | 25 | 25 |
| Bunch length [mm] | 75.5 | 15.8 |
| Beam-beam parameter, $\xi_{pp}$ | 0.006 | 0.008 |

Using Equation 2, one obtains μp collider parameters given in Table 13 (see [27] for details).

**Table 13.** Main parameters of the SppC based μp colliders.

| $E_\mu$ [TeV] | $E_p$ [TeV] | $\sqrt{s}$ [TeV] | $L_{\mu p}$ [$10^{32}$ cm$^{-2}$s$^{-1}$] | $\xi_\mu$ [$10^{-3}$] | $\xi_p$ [$10^{-2}$] |
|---|---|---|---|---|---|
| 0.75 | 35.6 | 10.33 | 5.5 | 8.7 | 6.0 |
| 0.75 | 68 | 14.28 | 12.5 | 8.7 | 8.0 |
| 1.5 | 35.6 | 14.61 | 4.9 | 8.7 | 6.0 |
| 1.5 | 68 | 20.2 | 42.8 | 8.7 | 8.0 |

Implementing parameters from Tables 9 and 12 into AloHEP, we obtain values of luminosity and beam-beam tune shifts given in Table 14.

**Table 14.** AloHEP results for the SppC based *μp* colliders

| $E_\mu$ [TeV] | $E_p$ [TeV] | $\sqrt{s}$ [TeV] | $L_{\mu p}$ [$10^{32}$ cm$^{-2}$s$^{-1}$] | $\xi_\mu$ [$10^{-3}$] | $\xi_p$ [$10^{-2}$] |
|---|---|---|---|---|---|
| 0.75 | 35.6 | 10.33 | 6.9 | 8.7 | 6.0 |
| 0.75 | 68 | 14.28 | 16 | 8.7 | 8.0 |
| 1.5 | 35.6 | 14.61 | 6.2 | 8.7 | 6.0 |
| 1.5 | 68 | 20.2 | 50 | 8.7 | 8.0 |

## 2.8. U. Kaya et al., "The LHC based μp colliders"

Parameters of the LHC proton beam and MC muon beam are presented in Tables 15 and 16, respectively.

**Table 15.** Main parameters of proton beam

| Parameter | HL-LHC |
|---|---|
| Proton energy [TeV] | 7 |
| Bunch intensity [$10^{11}$] | 2.2 |
| Normalized emittance [μm] | 2.0 |
| Bunch spacing [ns] | 25 |
| Number of bunches per ring | 2760 |
| $\beta^*$ possible for e-p at IP [m] | 0.07 |
| Beam size at IP [μm] | 4.45 |

**Table 16.** Muon collider design parameters.

| Parameter | | | |
|---|---|---|---|
| Muon energy [TeV] | 0.75 | 1.5 | 3 |
| Repetition rate [Hz] | 15 | 12 | 6 |
| L [$10^{34}$ cm$^{-2}$s$^{-1}$] | 1.25 | 4.6 | 11 |
| Circumference [km] | 2.5 | 4.4 | 6 |
| $\beta^*$ at IP [cm] | 1 | 0.5 | 0.3 |
| $\varepsilon_N$ [π mm·mrad] | 25 | 25 | 25 |
| # per bunch [$10^{12}$] | 2 | 2 | 2 |
| Bunches per beam | 1 | 1 | 1 |
| $\xi_\mu$ at IP | 0.09 | 0.09 | 0.09 |
| $\sigma_\mu$ at IP [μm] | 6 | 3 | 1.64 |

Using Equation 2, one obtains *μp* collider parameters given in Table 17 (see [30] for details).

**Table 17.** Main parameters of the LHC based *μp* collider.

| Muon Energy [TeV] | 0.75 | 1.5 | 3 |
|---|---|---|---|
| $\sqrt{s}$ [TeV] | 4.58 | 6.48 | 9.16 |
| L [$10^{33}$ cm$^{-2}$s$^{-1}$] | 1.4 | 2.3 | 0.9 |

Implementing parameters from Tables 15 and 16 into AloHEP, we obtain values of luminosity and beam-beam tune shifts given in Table 18.

**Table 18.** AloHEP results for the LHC based *μp* colliders

| $E_\mu$ [TeV] | $L_{\mu p}$ [$10^{33}$ cm$^{-2}$s$^{-1}$] | $\xi_\mu$ [$10^{-3}$] | $\xi_p$ [$10^{-2}$] |
|---|---|---|---|
| 0.75 | 1.7 | 9.5 | 12 |
| 1.5 | 2.9 | 9.5 | 12 |
| 3.0 | 2.2 | 9.5 | 12 |

## 2.9. U. Kaya et al., "Luminosity and Physics Considerations on HL-LHC and HE-LHC based mu-p Colliders"

HL-LHC and HE-LHC based *μp* colliders were proposed in Reference [31]. Possible layout is shown in Figure 5. Nominal parameters of proton beams are presented in Table 19, parameters of the proton ring upgraded for ERL60 related ep colliders are given in Table 20 (for details see ref [31]). Parameters of muon beams are given in Table 21 (which is slightly different from Table 16).

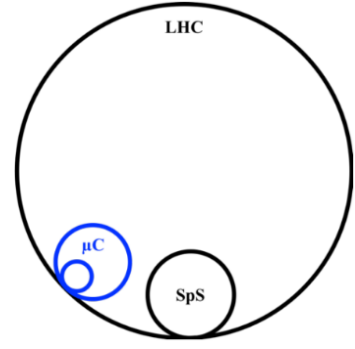

**Figure 5.** Schematic drawing of proposed LHC based *μp* colliders.

**Table 19.** HL-LHC and HE-LHC design parameters

| Parameter [Unit] | HL-LHC | HE-LHC |
|---|---|---|
| $\sqrt{s}$ [TeV] | 14 | 27 |
| Circumference [km] | 26.7 | 26.7 |
| Beam current [A] | 1.12 | 1.12 |
| Bunch population [$10^{11}$] | 2.2 | 2.2 |
| Bunches per beam | 2760 | 2808 |
| Bunch length [mm] | 90 | 90 |
| Bunch spacing [ns] | 25 | 25 |
| $\varepsilon_N$ [mm·mrad] | 2.5 | 2.5 |
| $\beta^*$ function at IP [m] | 0.15 | 0.45 |
| Beam size at IP [μm] | 7.1 | 9 |
| L [$10^{34}$ cm$^{-2}$s$^{-1}$] | 5 | 16 |

**Table 20.** ERL60 ⊗ (HL-LHC & HE-LHC) *ep* collider parameters

| Parameter [Unit] | HL-LHC | HE-LHC |
|---|---|---|
| $E_p$ [TeV] | 7 | 13.5 |
| $E_e$ [TeV] | 60 | 60 |
| $\sqrt{s}$ [TeV] | 1.3 | 1.7 |
| Bunch spacing [ns] | 25 | 25 |
| Protons per bunch [$10^{11}$] | 2.2 | 2.5 |
| Proton $\varepsilon_N$ [μm] | 2 | 2.5 |
| Electrons per bunch [$10^9$] | 2.3 | 3 |
| Electron current [mA] | 15 | 20 |
| Circumference [km] | 26.7 | 26.7 |
| Beam current [A] | 1.12 | 1.12 |
| Proton $\beta^*$ at IP [cm] | 7 | 10 |
| Proton beam size at IP [μm] | 4.45 | 4.17 |
| L [$10^{33}$ cm$^{-2}$s$^{-1}$] | 8 | 12 |

**Table 21.** Muon collider parameters.

| Parameter [Unit] | Multi-TeV | | |
|---|---|---|---|
| $\sqrt{s}$ [TeV] | 1.5 | 3.0 | 6.0 |
| Circumference [km] | 2.5 | 4.5 | 6 |
| Bunch population [$10^{12}$] | 2 | 2 | 2 |
| Bunch length [cm] | 1.0 | 0.5 | 0.2 |
| $\varepsilon_N$ [μm] | 25 | 25 | 25 |
| $\beta^*$ at IP [cm] | 1 | 0.5 | 0.25 |
| Beam size at IP [μm] | 5.9 | 2.95 | 1.48 |
| Repetition rate [Hz] | 15 | 12 | 6 |
| Number of IPs | 2 | 2 | 2 |
| L [$10^{34}$ cm$^{-2}$s$^{-1}$] | 1.25 | 4.4 | 12 |

Using the nominal (upgraded) parameters of HL-LHC and MC from Tables 19, 20 and 21, the parameters of HL-LHC based *μp* colliders estimated according to Equation (2) are presented in Table 22.

**Table 22.** Center of mass energies and luminosities of HL-LHC based μp colliders.

| $E_\mu$ [TeV] | $\sqrt{s}$ [TeV] | L (nominal) [$10^{33}$ cm$^{-2}$s$^{-1}$] | L (upgraded) [$10^{33}$ cm$^{-2}$s$^{-1}$] |
|---|---|---|---|
| 0.75 | 4.58 | 0.95 | 1.4 |
| 1.5 | 6.48 | 0.84 | 2.1 |
| 3 | 9.16 | 0.57 | 1.5 |

Center-of-mass energies and luminosity values for HE-LHC based *μp* colliders, evaluated in the same way, are given in Table 23.

**Table 23.** Center of mass energies and luminosities of HE-LHC based μp colliders.

| $E_\mu$ [TeV] | $\sqrt{s}$ [TeV] | L (nominal) [$10^{33}$ cm$^{-2}$s$^{-1}$] | L (upgraded) [$10^{33}$ cm$^{-2}$s$^{-1}$] |
|---|---|---|---|
| 0.75 | 6.36 | 0.59 | 1.6 |
| 1.5 | 9 | 0.52 | 2.8 |
| 3 | 12.7 | 0.36 | 1.9 |

Matching transverse sizes of proton and muon beams in Equations (3) and (4), $\xi_p$ and $\xi_\mu$ turn into:

$$\xi_p = \frac{N_\mu r_p}{4\pi \varepsilon_N^p} \quad (5)$$

$$\xi_\mu = \frac{N_p r_\mu}{4\pi \varepsilon_N^\mu} \quad (6)$$

Putting the corresponding parameters from Tables 19–21 into Equations (5) and (6), we obtain the tune shift values given in Table 24.

**Table 24.** Beam-beam tune shifts.

| | | $\xi_p$ | $\xi_\mu$ |
|---|---|---|---|
| HL-LHC | Nominal | 0.098 | 0.0096 |
| | Upgraded | 0.12 | 0.0096 |
| HE-LHC | Nominal | 0.098 | 0.0096 |
| | Upgraded | 0.098 | 0.011 |

Using AloHEP, we have double-checked the parameters given in Tables 22-24, resulting in a well-consistent outcome.

**2.10. K. Cheung et al.**, "Physics potential of a muon-proton collider"

Physics search potential of the LHC and FCC based *μp* colliders has been considered in a recent paper [28]. Basic parameters of corresponding colliders are presented in Table 25.

**Table 25.** Basic parameters of the two μp experiments.

| Exp. | $E_p$ [TeV] | $E_\mu$ [TeV] | $\sqrt{s}$ [TeV] | $L_{int}$ [ab$^{-1}$] |
|---|---|---|---|---|
| μp - 1 | 7 | 1 | 5.3 | 0.1 |
| μp - 2 | 50 | 3 | 24.5 | 1 |

Since muon and proton beam parameters are not given in Reference [28], it is not possible to check luminosity values by AloHEP. The first option is comparable with Table 22 assuming 10 years operation. Integrated luminosity for second option seems to be over estimated by an order (see Table 8). However, luminosity for this option may be upgraded in case of a dedicated design of proton beam for *μp* collisions (values in Table 8 are estimated for synchronous operation of *μp* and *pp* colliders).

**2.11. D. Acosta et al.**, "A Muon-Ion Collider at BNL: the future QCD frontier and path to a new energy frontier of μ$^+$μ$^-$ colliders"

Recently, Muon-Ion Collider at BNL has been proposed in Reference [29] as a demonstrator toward a future muon–antimuon collider at 10 TeV energy. Here, both proton and gold nucleus options have been considered keeping huge physics search potential of these colliders in mind. As it is mentioned in Introduction, only *μp* colliders have been reviewed in this paper, whereas *μA* colliders will be reviewed in forthcoming paper.

Main parameters of the MuIC at BNL are summarized in Table 26 (Table 2 of Reference [29]).

**Table 26.** The proposed parameters of the MuIC at BNL.

| Parameter | Muon | Proton |
|---|---|---|
| Energy [TeV] | 0.96 | 0.275 |
| CoM energy [TeV] | 1.03 | |
| Bunch intensity [$10^{11}$] | 20 | 3 |
| Norm. Emittance, $\varepsilon_{x,y}$ [μm] | 25 | 0.2 |
| $\beta^*$ at IP [cm] | 1 | 5 |
| Trans. Beam size, $\sigma_{x,y}$ [μm] | 5.2 | 5.8 |
| Muon repetition rate, $f_{rep}$ (Hz) | 15 | |
| Cycles/ collisions per muon bunch, $N_c$ | 3279 | |
| $L_{\mu p}$ [$10^{33}$ cm$^{-2}$s$^{-1}$] | 7 | |

Implementing parameters from Table 26 into AloHEP, we obtain L=6.9×$10^{33}$cm$^{-2}$s$^{-1}$ without muon decays and L=4.35×$10^{33}$cm$^{-2}$s$^{-1}$ with muon decays.

**5. Conclusion**

Construction of future muon collider (or dedicated *μ*-ring)

tangential to existing and proposed hadron colliders will give opportunity to realize µp colliders with multi-TeV center-of-mass energies at a luminosity of order of $10^{33}$ cm$^{-2}$s$^{-1}$ ($10^{34}$ cm$^{-2}$s$^{-1}$). Obviously, such colliders will essentially enlarge the physics search potential of corresponding muon and hadron colliders for both the SM and BSM phenomena. Therefore, systematic studies of accelerator, detector and physics search aspects of the LHC/FCC/SppC based µp colliders are necessary for long-term planning of High Energy Physics.

Concerning SM physics (especially QCD), search potential of µp colliders is approximately similar with linac-ring type ep colliders with similar center-of-mass energies. This statement is valid for BSM physics in general. However, µp colliders are unique in search of muon-related new physics. In Figure 6, production of color-octet muons is presented as an example (see Reference [31] for details).

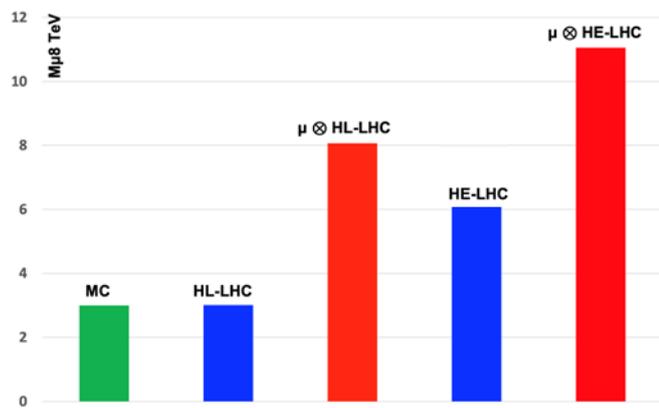

**Figure 6.** Discovery limits for color octet muon at the MC with √s=6 TeV, HL-LHC, HE-LHC and corresponding µp colliders

Keeping huge physics search potential of µp colliders in mind, we propose establishment of an international working group for systematic studies of accelerator, detector and physics search aspects of the LHC/FCC/SppC based µp colliders.

**Acknowledgements**

Authors are grateful to Ahmet Nuri Akay, Ali Can Canbay, Hande Karadeniz, Umit Kaya, Bilgehan Baris Oner, Arif Ozturk and Frank Zimmermann for useful cooperation.